\documentclass[12pt]{iopart}
\usepackage[dvips]{graphicx}

\begin{document}

\title{Spreading dynamics in complex networks}

\author{Sen Pei$^{1,2}$ and Hern\'an A. Makse$^{2*}$}
\address{$^1$LMIB and School of Mathematics and Systems
Science, Beihang University, Beijing, 100191, China}
\address{$^2$Levich Institute and Physics Department, City College of
New York, New York, NY 10031, USA}
\ead{$^*$hmakse@lev.ccny.cuny.edu.}
\begin{abstract}

Searching for influential spreaders in complex networks is an issue
of great significance for applications across various domains,
ranging from the epidemic control, innovation diffusion, viral
marketing, social movement to idea propagation. In this paper, we
first display some of the most important theoretical models that
describe spreading processes, and then discuss the problem of
locating both the individual and multiple influential spreaders
respectively. Recent approaches in these two topics are presented.
For the identification of privileged single spreaders, we summarize
several widely used centralities, such as degree, betweenness
centrality, PageRank, $k$-shell, etc. We investigate the empirical
diffusion data in a large scale online social community --
LiveJournal. With this extensive dataset, we find that various
measures can convey very distinct information of nodes. Of all the
users in LiveJournal social network, only a small fraction of them
involve in spreading. For the spreading processes in LiveJournal,
while degree can locate nodes participating in information diffusion
with higher probability, $k$-shell is more effective in finding
nodes with large influence. Our results should provide useful
information for designing efficient spreading strategies in reality.

\end{abstract}

%Uncomment for PACS numbers title message
%\pacs{00.00, 20.00, 42.10}
% Keywords required only for MST, PB, PMB, PM, JOA, JOB?
\vspace{2pc}
\noindent{\it Keywords}: spreading process, influential
spreaders, complex networks

% Uncomment for Submitted to journal title message
%\submitto{\JSTAT}
% Comment out if separate title page not required
\maketitle

\section{Introduction}

As a widespread process, spreading describes many important
activities in real world, ranging over the outbreak of epidemics
\cite{Anderson1992,Hethcote2000,Diekmann2000,Keeling2008}, the
spread of news and ideas
\cite{Adar2004,Kumar2004,Adar2005,Leskovec2009}, the diffusion of
technique innovations \cite{Metcalfe1987,Strang1998,Rogers2010}, the
promotion of commercial products
\cite{Richardson2002,Domingos2005,Watts2007,Leskovec2007}, and the
rise of political movements
\cite{Diani2003,McAdam1993,Polletta2001,Snow1980}. Understanding the
mechanism behind the global spread of an epidemic or information is
fundamental for applications in a variety of fields as diverse as
epidemiology \cite{Diekmann2000,Keeling2008}, viral marketing
\cite{Watts2007,Leskovec2007}, collective dynamics
\cite{Albert2002,Boccaletti2006,Barrat2008} and robustness of
networks \cite{Albert2000,Newman2003,Cohen2001}. In real life, the
diffusion of a contagious disease or a piece of information usually
underlies individuals' contacts. Take the influenza as an example.
The transmission of the influenza virus mainly depends on the direct
contacts of infected persons and susceptible people. As people's
interaction is responsible for these spreading processes, the
position of a person in the complex network, which is formed
according to individuals' social relations, usually determines the
spreading capability of this person. As the empirical research has
shown, the complex networks in reality, which describe various
systems in the fields of social science
\cite{Wasserman1994,Freeman2004,Castellano2009,Gallos2012,Yan2013},
neuroscience \cite{Kinouchi2008,Larremore2011,Pei2012}, ecology
\cite{Camacho2002,Dunne2002} and economics
\cite{Yakovenko2009,Pei2012a}, are by no means randomly connected.
On the contrary, they display many non-trivial topological features
such as the heavy-tailed degree distribution \cite{Barabasi1999},
small world effect \cite{Watts1998}, high clustering coefficient
\cite{Strogatz2001}, self-similarity \cite{Song2005,Gallos2012a} and
community structures \cite{Girvan2002,Galvao2010}. Therefore, people
in social networks with such rich topological structures should vary
a lot in spreading abilities.

For the people in social networks, a few influential spreaders are
able to start large scale diffusions and thus become more important
than the other persons in the spreading process \cite{Kitsak2010}.
How to identify these privileged nodes is of great significance for
applications across various domains. For instance, the knowledge of
influential spreaders is crucial for the design of efficient
strategies to control the outbreak of epidemics
\cite{Satorras2001,Boguna2002}. And targeting the vital people in
information dissemination is helpful for conducting successful
campaigns in commercial product promotions
\cite{Watts2007,Leskovec2007}. Due to its utmost importance in
practice, the problem of searching for influential spreaders in
complex networks has attracted much attention.

In the issue of identifying super spreaders, there are two distinct
subtopics: searching for individual influential spreaders
\cite{Kitsak2010} and a set of nodes that can maximize the influence
\cite{Richardson2002}. In the epidemic outbreaks or rumor diffusion,
the spreading usually starts from a single source node. Therefore,
monitoring the influential individual originators in social networks
is important in preventing or decelerating the spreading. For
locating individual spreaders, usually the topological or dynamical
measures are utilized. We first rank the nodes according to specific
measures and then select the nodes ranking top as influential
spreaders. In the field of viral marketing, however, it is usually a
set of nodes that are selected to start the propagation. The choice
of the multiple originators should maximize the final influence
since the goal of viral marketing is to persuade more people to buy
the commercial products with less cost. A trivial way to find the
best multiple spreaders is to select the nodes which are influential
as individual spreaders. However this attempt usually fails for the
reason that the top individual nodes tend to have large overlap in
their infected population \cite{Kitsak2010}. In fact, the issue of
finding a $k$-set nodes that can lead to maximal influence is a
NP-hard optimization problem \cite{Kempe2003}. The solution mainly
relies on the heuristic algorithms based on dynamic models in
complex networks.

While the overall topic of spreading is too wide even to introduce
briefly, we will mainly focus on the two subtopics described above.
In this paper, we first introduce some of the most important
theoretical models for spreading, which are widely used in the
research of influential spreaders. Then the measures for identifying
individual influential spreaders are discussed. We discuss their
features, calculation complexity and performance in locating best
single spreaders. After that, we present some progresses in finding
multiple influential spreaders with heuristic algorithms. Finally,
we show the empirical results based on extensive data collected from
large-scale online communities.

\section{Theoretical models for spreading}

The early approach of designing mathematical models that describe
spreading processes starts in the fields of sociology and
epidemiology
\cite{Schelling1978,Granovetter1978,Granovetter1983,Herrmann1986}.
Formulating theoretical models that capture real diffusion is
helpful for us to understand how a disease or information spreads
across a population. More importantly, the spreading models can be
used to predict the outcome of spreading, thus providing
instructions to accelerate or decelerate the diffusion processes.
After the initial works, many models are proposed in the
applications of a series of social and biological phenomena.

In the spreading models, there are usually two important elements to
be specified. The first one is the underlying network, which
describes how individuals interact with one another. The network is
recorded by a graph $G(N,E)$ with $N$ nodes and $E$ edges. If node
$i$ and $j$ have a chance to contact in reality, then there exists a
social link between them. As has been shown, the topological
structure of the social network can dramatically affect the outcome
of spreading \cite{Satorras2001,Boguna2003}. Another element that is
vital for models is the spreading rule by which information or
disease diffuses from one person to another. According to the
diffusion strategy, existing models of spreading typically fall into
two categories: independent interaction models and threshold models.
We will introduce these two types of models in details in this
section and explain some of their features.

\subsection{Independent interaction models}

In the epidemic spreading, each time an infected node contacts with
a susceptible node, there is a chance that the susceptible person
gets infected. Based on this fact, independent interaction models
assume that each interaction results in contagion with independent
probability. Specifically, whenever a susceptible person $j$ exposes
to an infected person $i$, $j$ will get infected with a probability
$p_{ij}$, which is not affected by the contacts with other nodes.
With such dynamics, independent interaction models mimic the
contagious process directly and imply the fact that spreading
underlies peoples's interaction. The more a susceptible individual
contacts with infected people, the higher probability he will be
infected. The susceptible-infected-recovered (SIR) and
susceptible-infected-susceptible (SIS) models from epidemiology
\cite{Anderson1992,Hethcote2000,Diekmann2000}, and the Bass model
from the innovation diffusion literature \cite{Bass1969} are
examples of independent interaction models. Here we mainly discuss
the well-known SIR and SIS models. Different models and
generalizations can be found in the references
\cite{Anderson1992,Hethcote2000,Diekmann2000,Keeling2008,Herrmann1986}.

As classical mathematical models for epidemic spreading, the SIR and
SIS models were first proposed by epidemiologists
\cite{Schelling1978,Granovetter1978,Granovetter1983,Herrmann1986}.
Since the mechanisms of SIR and SIS models are suitable to describe
various spreading processes, such as disease spreading, idea
propagation, and innovation diffusion, they have long been used in
the study of spreading. In SIR model, each person can be in one of
three possible states \cite{Anderson1992}, susceptible (S), infected
(I), or recovered (R). Susceptible individuals are healthy persons
that may catch the disease. Infected people stand for the persons
who have got the disease and are able to spread it to susceptible
individuals. A person is recovered if he has been cured from the
disease and becomes immune to it. In the classical SIR model, there
are two adjustable parameters: the transmission probability
$\lambda$ and the recovery probability $\mu$. The states of
individuals evolve as follows:
\begin{equation}
S(i)+I(j)\rightarrow _{\lambda} I(i)+I(j),\label{SI}
\end{equation}
\begin{equation}
I(i)\rightarrow _{\mu} R(i),\label{IR}
\end{equation}
where $i$ and $j$ are two neighbors in the social network. While the
SIR model is suitable to describe the spreading of disease with
immunity, there are many epidemics that an individual can catch for
more than once. In this case, the SIS model, in which we only
consider the susceptible (S) and infected (I) states, can better
describe the contagion. The contagion process of the SIS model is
the same as Eq.(\ref{SI}), while Eq.(\ref{IR}) is replaced by
$I(i)\rightarrow _{\mu} S(i)$. Generalizations of SIR and SIS models
can be implemented by imposing a distribution of the transmission
rate $\lambda$ and the recovery rate $\mu$.

According to the distinct dynamics, there are fundamental difference
in the outcomes of SIR and SIS models. In the SIR model, the
infection will eventually die out, because once an individual
becomes immune to the disease, he/she will never get infected again.
Whereas, for SIS models, people can be infected for many times. So
the disease can reach an endemic state, where a certain fraction of
population are kept infected. Considering this difference, when we
measure the result of SIR model, we are interested in the fraction
of individuals who have ever caught the disease. While for SIS
model, it is the fraction of infected nodes persisting in the
endemic state that we concern.

Besides the spreading strategy, it is meaningful to discuss the
impact of the underlying social network structure on the spreading
results. In the early research, both SIR and SIS models are
considered within the homogeneous mixing hypothesis
\cite{Anderson1992}, where the infectious and susceptible people
contact with each other randomly. In this case, the underlying
social network is actually an Erd\'{o}s-R\'{e}nyi random graph. The
most important observation under this condition is the emergence of
an epidemic threshold \cite{Murray1993}. Take the SIR model as an
example (without loss of generality we set $\mu=1$). When a single
node becomes infectious among susceptible population, the epidemic
threshold is given by $\lambda_c=1/\langle k\rangle$, where $\langle
k\rangle$ is the average number of connections of nodes. If
$\lambda>\lambda_c$, the disease will infect a finite fraction of
the population. On the other hand, if $\lambda<\lambda_c$, the
fraction of infected individual will tend to zero in the limit of
very large population.

As we all know, the transmission networks in real life are by no
means totally random. For example, the degree distribution of the
sexual contact network is found to be power-law \cite{Liljeros2001}.
To this end, several important works have been done to understand
the effects of the nontrivial network structure on the spreading
outcomes
\cite{Satorras2001,Grassberger1983,Moore2000,Newman2002,Satorras2001a,Satorras2002,Satorras2002a}.
For SIR model on uncorrelated graphs with a generic degree
distribution $P(k)$ and a finite average degree $\langle k\rangle$,
the epidemic threshold is defined by $\lambda_c=\langle
k\rangle/(\langle k^2\rangle-\langle k\rangle)$
\cite{Satorras2001,Newman2002,Satorras2001a,Moreno2002,Dezso2002}.
Apparently, for networks with $\langle k^2\rangle<\infty$, the
threshold has a finite value. Whereas, for networks with strongly
fluctuating degree distribution, the infinity of $\langle
k^2\rangle$ results in a vanishing epidemic threshold for large
scale networks. Analogously, the absence of an epidemic threshold in
scale-free networks with the power-law exponent $2<\gamma\leq3$ was
reported as well \cite{Satorras2001,Satorras2001a}. Apart from the
topological structures mentioned above, other complex topologies
have been considered, such as the high-clustering
\cite{Eguiluz2002,Serrano2006}, small-world
\cite{Moore2000,Kuperman2001,Santos2005}, degree correlation
\cite{Boguna2003,Moreno2003}, etc. These works provide more insights
into the interplay between network structure and epidemic outcomes
for SIR and SIS models.

\subsection{Threshold models}

Although the independent interaction models can describe the
epidemic spreading in reality, numerous phenomena in economics and
sociology
\cite{Granovetter1978,Gladwell2000,Bik1992,Aguirre1988,Glaeser1996,Valente1995,Schelling1973,Glance1993}
are better described by the threshold models. In these applications,
people tend to adopt a new behavior or information only if a certain
fraction of their neighbors have already done so
\cite{Watts2002,Kleinberg2007}. In this case, the effect of a single
interaction is no longer independent, but strongly depends on other
exposures.

The simplest threshold model is the Linear Threshold Model. In this
model, each node is assigned a threshold value, which is the
fraction of neighbors required for it to adopt the new behavior or
information. And on each link $(i,j)$, we define a weight to reflect
the influence that $j$ exerts on $i$. In the spreading process, some
initial nodes $S$ starts out adopting the new behavior. A node is
defined as active if it is following the new behavior. At a given
time, any inactive node becomes active if the sum of weights from
its active neighbors exceeds its threshold. In subsequent times, the
activation of some nodes may cause other nodes to adopt the new
behavior, and such process can be applied repeatedly. This
phenomenon is usually used to explain the cascading behavior in
social science \cite{Watts2002,Kleinberg2007}. More generally, based
on the Linear Threshold Model, we can assign each node a threshold
function instead of using the weighted sum in the state updating.
Such General Threshold Model is more general since it can reflect
any types of threshold rule.

Compared with independent interaction model, the threshold model
actually incorporates the memory of past exposures history.
Therefore, the result of a single interaction is determined by other
interactions. This radical difference turns out to have significant
impact on the spreading dynamics. With such threshold model, global
cascading which is triggered by small number of initial originators
is observed \cite{Watts2002}. Also it has been shown in a model with
memory of exposures history, the final state of the spreading is
controlled by only two parameters: $P_1$ and $P_2$, which stand for
the probability that a node becomes infected due to one and two
contacts respectively \cite{Dodds2004}. This indicates that the
interplay of single contacts in threshold model can lead to
different dynamics with independent interaction models.

Apart from the models mentioned above, there are also other
variations describing spreading in complex networks, including the
standard rumor model proposed by Daley and Kendal
\cite{Daley1964,Zanette2001,Liu2003,Moreno2004,Moreno2004a}, the
voter model \cite{Krapivsky2003,Fontes2002,Castellano2003,Sood2005},
the strategic game models
\cite{Morris2000,Blume1993,Ellison1993,Young1998}, etc. All these
models help us understand the mechanism of spreading in various
domains and many profound results have been applied in reality.

\section{Searching for individual influential spreaders}

In order to find effective predictors for individual influential
spreaders, various measures are designed to rank the nodes according
to their statuses in spreading. Most of the proposed measures are
determined by nodes' topological features as well as an assuming
spreading mechanism. Here we will introduce some of the most
important predictors that are widely used to quantify nodes'
spreading ability.

In the context of social science, the topology of a social network
is represented by an adjacency matrix $A=\{a_{ij}\}_{N\times N}$,
where the element $a_{ij}>0$ if there exists a link from $j$ to $i$
and $a_{ij}=0$ otherwise. For an undirected network, $A$ is a
symmetric matrix with $a_{ij}=a_{ji}$. If the network is weighted,
the element $a_{ij}$ represents the weight of the link from $j$ to
$i$. Actually, the adjacency matrix $A$ fully describes the
topological structure of the social network. Once one has the
adjacency matrix, it is possible to calculate the following measures
for each node. Some of the measures only need the local information,
i.e. the properties of a node's neighbors, while some others require
the complete structure of the social network. Due to the large size
of modern social networks, the calculation of global measures
imposes great challenge in the research of social networks.

\subsection{Degree}

In real social networks, it is observed that while most of the
people have small or moderate number of connections to other
individuals, there are very few of hubs that maintain extremely
large number of social relations. Such phenomenon is described by a
power-law (or heavy-tailed) degree distribution. For an unweighted
graph, degree is the number of links connecting to a node. For a
weighted graph, it is defined as the sum of weights from edges
connecting to a node. In an undirected network, according to the
adjacency matrix of a graph, degree $k(i)$ for a node $i$ can be
computed as follows:
\begin{equation}
k(i)=\sum_{j=1}^{N}a_{ij}.\label{degree}
\end{equation}
In the case of a directed network, we usually define two separate
measures of degree centrality, namely indegree and outdegree.
Concretely, indegree is the number of links directed to the node and
outdegree is the number of links that the node directs to others.
Indegree and outdegree are defined as:
\begin{equation}
k_{in}(i)=\sum_{j=1}^{N}a_{ij}.\label{indegree}
\end{equation}
\begin{equation}
k_{out}(i)=\sum_{j=1}^{N}a_{ji}.\label{outdegree}
\end{equation}

Computing degree centrality for all the nodes in a dense network
takes complexity $O(V^2)$. However, for a network with a sparse
adjacency matrix, which we usually encounter in reality, its
computational complexity is reduced to $O(E)$, making the degree
centrality a feasible measure even for very large networks. Albeit
it is a local measure, degree is efficient in finding important
nodes in many situations. For example, for epidemic spreading in
scale-free networks, hubs are more likely to be infected and can
lead to large scale diffusion \cite{Satorras2001}. Also in complex
networks with broad degree distribution, such as Internet, power
grid or other infrastructure networks, intentional attack of the
hubs can result in rapidly breakdown of the whole structure
\cite{Albert2000,Cohen2001}. Such fragility of scale-free networks
under intentional attacks indicates that hubs play prominent role in
the structure stability. However, not all hubs are guaranteed to be
super spreaders. For instance, if a hub locates in the periphery of
the network, its spreading ability would be limited
\cite{Kitsak2010}. After all, degree only captures the number of the
nearest neighbors of a node. In fact, the spread capability of the
neighbors can also affect the nodes' importance in spreading
significantly. Therefore, measures involved with more information
are desired to improve the performance of degree. Even so, due to
its easy accessibility and relatively satisfied performance, degree
is still used as an effective predictor of influential spreaders in
many applications.

\subsection{Betweenness and closeness centralities}

Betweenness and closeness centrality are two well-known ranking
measures in social science \cite{Freeman1979,Friedkin1991}. Both of
them are proposed based on the assumption that information tend to
traverse the network from the originator to destination through the
shortest path.

In the social network theory, betweenness centrality is defined as a
measure of how many shortest paths cross through this node
\cite{Freeman1979,Friedkin1991}. For a network $G=(V,E)$, the
betweenness centrality of node $i$, denoted by $C_B(i)$ is defined
as
\begin{equation}
C_B(i)=\sum_{s\neq i\neq t\in V}\frac{\sigma_{st}(i)}{\sigma_{st}},
\label{betweenness}
\end{equation}
where $\sigma_{st}$ is the number of shortest paths between nodes
$s$ and $t$, and $\sigma_{st}(i)$ is the number of shortest paths
between $s$ and $t$ which pass through node $i$. With this
definition, the nodes with large betweenness centrality usually hold
the vital positions in the shortest pathways between large numbers
of pairs of nodes.

Apart from the betweenness centrality, in a connected network, there
exists a natural distance metric between all pairs of nodes. The
farness \cite{Sabidussi1966} of a node $s$ is defined as the sum of
its distances to all other nodes, and its closeness centrality is
defined as the inverse of the farness. Thus, the smaller closeness a
node has, the lower its total distance to all other nodes.
Precisely, closeness centrality of node $i$ is defined as
\cite{Sabidussi1966}:
\begin{equation}
C_C(i)=\frac{1}{\sum_{t\in V\setminus i}d_{G}(i,t)},
\label{closeness}
\end{equation}
where $d_G(i,t)$ is the shortest distance between $i$ and $t$. In
fact, closeness centrality can be viewed as a measure of how long it
will take for a piece of information to spread from a given node to
other reachable nodes through the shortest paths in the network. The
smaller a node's closeness is, the faster the information diffuses
from this node.

However, on an unconnected network, the closeness centrality is not
well defined. Since the distance between any two unreachable nodes
is infinity, the closeness centrality of all nodes in an unconnected
graph would be 0. To solve this problem, a modified version of the
classic closeness, residual closeness \cite{Dangalchev2006}, is
proposed. The residual closeness of node $i$ is defined as
\begin{equation}
C_R(i)=\sum_{t\in V\setminus i} 2^{-d_{G}(i,t)}.
\label{residualcloseness}
\end{equation}

In general, betweenness and closeness centralities can identify
crucial nodes in transportation. Take the betweenness centrality as
an example, the nodes with large betweenness usually hold the vital
positions in the pathways between pairs of nodes. If such nodes are
intentionally attacked, the overall efficiency of spreading will be
heavily damaged, since the increase of path length would make it
difficult for a piece of information to spread to other nodes. In
the networks with heavy-tailed degree distribution, hubs usually
serve as intermediate nodes in the shortest paths between nodes
\cite{Boguna2008}. So hubs incline to have large betweenness
centrality. Besides, the nodes connecting two separate communities
also have large betweenness centrality. Such nodes, although not
necessarily being well connected, play the role of connecting
bridges in the transportation between the nodes in two communities.
Applications of betweenness centrality include computer and social
networks \cite{Brandes2008}, biology
\cite{Estrada2007,Martin2010,Zuo2012}, transport \cite{Wang2011},
scientific cooperation \cite{Abbasi2012} and so forth.

Comparing with degree, betweenness and closeness centralities care
more about the global structure. Therefore, the calculation of
betweenness and closeness requires the complete network structure.
Moreover, both betweenness and closeness centralities involve
calculating the shortest paths between all pairs of vertices on a
graph, which is a rather time-consuming task. The classic algorithm
finding shortest paths between all pairs of nodes is the
Floyd-Warshall algorithm \cite{Floyd1962}, and it will take the
complexity $O(V^3)$. Later on, some more efficient algorithms are
developed for specific types of networks. On sparse networks,
Johnson algorithm requires $O(V^2\log V+VE)$ time to compute the
betweenness centrality \cite{Johnson1977}. Brandes has proposed a
more efficient algorithm. For unweighted sparse networks, it has the
complexity of $O(VE)$ \cite{Brandes2001}. Even with the fastest
algorithm, for large scale online social networks with tens of
millions of nodes, such as Twitter and Facebook, it is usually
infeasible to get the betweenness or closeness centralities in a
reasonable time.

\subsection{Eigenvector and PageRank centralities}

Eigenvector centrality was first introduced in the research of
sociology \cite{Bonacich1972}, where it was used to measure the
influence of a person in a social network. The main idea behind
eigenvector centrality is that a node's importance is not only
determined by itself, but also affected by its neighbors'
importance. A node connecting to important nodes will make itself
also important. With this idea, the eigenvector centrality of vertex
$i$ can be defined as:
\begin{equation}
e(i)=\frac{1}{\lambda}\sum_{j=1}^{N}a_{ij}e(j), \label{eigenvector}
\end{equation}
where $\lambda$ is a constant and $a_{ij}$ is the entry of the
adjacency matrix $A=\{a_{ij}\}_{N\times N}$. Actually, this equation
can be rewritten in vector notation as
\begin{equation}
\mathbf{Ae}=\lambda\mathbf{e}. \label{eigenvectormatrix}
\end{equation}
In the matrix theory, there will be many different eigenvalues
$\lambda$ for which an eigenvector solution exists. However, when we
quantify the influence of a node, it is required that the measure
should be positive. According to the Perron-Frobenius theorem, only
the largest eigenvalue can lead to such centrality measure
\cite{Newman2008}. Clearly, eigenvector centrality not only depends
on the degree of the nodes, but also on their neighbors' eigenvector
centrality. Due to this recursive property, eigenvector centrality
can reflect the global feature of the network.

As a generalization and variation of eigenvector centrality,
PageRank was originally introduced to rank web pages in the world
wide web (www) \cite{Brin1998}. As a successful ranking algorithm,
it is not only adopted by webpage search engines like Google, but
also used in ranking the importance of elements in a wide range of
applications, such as scientific ranking
\cite{Chen2007,Walker2007,Schmidt2007}, gene research
\cite{Chen2009,Ivan2011}, traffic and transportation
\cite{Jiang2008}, ecological systems \cite{Allesina2009}, and even
lexical semantics \cite{Navigli2010}. Compared with eigenvector
centrality, PageRank introduces a small probability of random
jumping to handle walking traps on a graph. The PageRank of a node
in a network can be calculated from
\begin{equation}
p_t(i)=\frac{1-\alpha}{N}+\alpha\sum_{j}\frac{a_{ij}p_{t-1}(j)}{k_{out}(j)},
\label{PageRank}
\end{equation}
where $k_{out}(j)$ is the number of outgoing links from node $j$ and
$\alpha$ is the jumping probability. Equation (\ref{PageRank})
actually describes a random walk process: a random walker moves
along the links of the network with probability $\alpha$, and jumps
to a randomly selected node with probability $1-\alpha$. $p_t(i)$ is
the probability that node $i$ is visited by the random walker at
time $t$. As time $t$ increases, the probability $p_t(i)$ will
converge to a stationary probability $p(i)$. This value is defined
as the PageRank which are used to determine its ranking relative to
other nodes. In the calculation, the conventional choice of $\alpha$
is 0.85. Different choice of $\alpha$ can affect the ranking
results.

From a calculation aspect, both eigenvector and PageRank
centralities can be computed efficiently by power iteration
\cite{Arasu2002,Franceschet2011}. Initially assign each node with
the same score, and then iterate according to corresponding update
equations. The result usually converges quickly in iterations. So
eigenvector and PageRank centralities can be applied to large scale
networks.

Based on the classic PageRank algorithm, several variations are
proposed. One is the so-called LeaderRank \cite{Lu2011}. On the
basis of PageRank, LeaderRank introduces a ground node $g$, which
has two directed links to every node in the original network. In
this way, the network will become strongly connected. More
importantly, LeaderRank is a parameter-free algorithm, thus getting
rid of the influence of parameters. Although LeaderRank stems from
PageRank, it is reported to be more stable to noisy data containing
spurious and missing links. In the condition where spammers create
fake links to obtain high rank, LeaderRank performs more reliable
than PageRank in ranking users. Also an extension of PageRank
algorithm, the TwitterRank \cite{Weng2010}, is proposed to measure
the influence of users in Twitter. TwitterRank measures the
influence taking both the topical similarity between users and the
link structure into consideration. Experimental results show that
TwitterRank outperforms the one Twitter currently uses and other
related algorithms, including the classical PageRank and
Topic-sensitive PageRank \cite{Haveliwala2002}.

\subsection{$k$-shell index}

In the graph theory, $k$-shell index describes the location of a
person in the social network
\cite{Bollobas1984,Seidman1983,Carmi2007}. The $k$-shell index of a
node is obtained by a procedure called $k$-shell decomposition,
where we successively prune nodes in the network layer by layer.
Concretely, the decomposition starts by removing nodes with degree
$k=1$. After that, some nodes may have only one link left. So we
continue pruning the network iteratively until there are no nodes
with $k=1$. The removed nodes fall into a $k$-shell with index
$k_S=1$. With the similar method, we iteratively remove the next $k$
shell $k_S=2$ and higher $k$ shells until all nodes are pruned. In
the decomposition procedure, each node is assigned with a $k$-shell
index. The periphery of the network corresponds to small $k_S$ and
the nodes with high $k_S$ define the core of the network.

Compared with the degree $k$, the $k_S$ index provides a different
type of information. By definition, a given layer with index $k_S$
can be occupied with nodes of degree $k\geq k_S$. For random model
networks, a strong correlation between $k$ and the $k_S$ index of a
node exists: the nodes with lower degree incline to stay in the
periphery of network while the core region is mainly occupied by
hubs. Therefore, both the degree and $k$-shell provide similar
information. In real networks, however, this relation is often not
true. In real networks hubs may have very different $k_S$ values and
can be located both in the periphery or in the core of the network
\cite{Kitsak2010}.

Recently, it is reported that for SIR and SIS modeling, the most
influential spreaders, which can lead to large scale epidemics, are
located in the inner core of the network (large $k_S$ region)
\cite{Kitsak2010}. The authors perform SIR and SIS models on a
series of real networks, and find that the spreading processes
originate from high $k$-shell nodes have larger average infected
population than those starting from nodes with high degree and
betweenness centrality. These results indicate that the $k$-shell
index of a node is a better predictor of spreading influence than
the commonly adopted degree and betweenness centrality. When a
spreading process starts in the core of the network, the epidemic or
information can diffuse through many pathways to the rest part of
the network. Moreover, it has been shown that nodes with high $k_S$
are easier to be infected and will be infected earlier than other
nodes. The nodes located in the core region of network tend to have
well connected neighbors, and the neighbors of their neighbors are
also prone to have large degree. It is these well-connected
neighbors that make the nodes in the core region more efficient in
spreading. Clearly, $k$-shell decomposition requires the complete
network structure. Once we have the adjacency matrix, $k$-shell
decomposition can be performed with complexity $O(E)$
\cite{Batagelj2003}. Therefore, the $k$-shell index can be applied
to large scale networks.

Despite its effectiveness, $k$-shell also have several defects. In
some situations, $k$-shell is limited due to the lack of resolution.
For example, in the networks with a tree structure, or networks
formed by growth models \cite{Barabasi1999}, $k$-shell index only
contains a few discrete values. Such degeneracy of the $k$-shell
index limits its predictive accuracy. This limitations have been
identified also in some empirical studies of spreading
\cite{Christakis2010,Castellano2012,Borge2012}.

Based on $k$-shell decomposition, several improvements or
alternatives are proposed. Zeng \cite{Zeng2013} proposed a mixed
degree decomposition (MDD) procedure in which both the residual
degree (number of links between the remaining nodes) and the
exhausted degree (number of links between the removed nodes) are
considered. In each step of the MDD procedure, the nodes are removed
according to the mixed degree $k_i^{(m)}=k_i^{(r)}+\lambda\times
k_{i}^{(e)}$, where $k_i^{(r)}$ is the residual degree, $k_i^{(e)}$
is the exhausted degree, and $\lambda$ is a tunable parameter
between 0 and 1. When $\lambda=0$, the MDD method becomes the
$k$-shell method, while when $\lambda=1$, the MDD method is
equivalent to the degree. By simulating the epidemic spreading
process on real networks, it is shown that the MDD method can
improve the performance of $k$-shell. Another method based on the
$k$-shell index is the $\mu$-power community index ($\mu$-PCI)
\cite{Basaras2013}, which is a balance of coreness and betweenness
centrality. The metric is computed as follows: the $\mu$-PCI of a
node $i$ is equal to $k$, such that there are up to $\mu\times k$
nodes in the $\mu$-hop neighborhood of $i$ with degree greater than
or equal to $k$, and the rest of the nodes in that neighborhood have
a degree less than or equal to $k$. Due to its computational
complexity, the authors only present the results for $\mu=1$. By
modeling on real networks, it is shown the 1-PCI exhibits steady and
reliable behavior. As 1-PCI values increase, influence also
continuously increases until maximum infection is reached.

\subsection{Path counting}

Path counting was first proposed as the accessibility metric
\cite{Travencolo2008,Viana2012}. The basic idea is to count the
number of all possible walks of arbitrarily length departing from
the source node. Recently, the concept of path counting was
exploited to quantify the importance of individual role in
collective dynamics \cite{Klemm2012}. A measure called dynamical
influence (DI) is proposed to quantify a node's influence in various
dynamical processes. The DI is calculated as the leading left
eigenvector of a characteristic matrix that records the topology and
dynamics. Specifically, in the characteristic matrix $M$, the entry
$M_{ij}$ stands for the influence that node $j$ exerts on node $i$.
The $i$th entry of the left eigenvector of $M$ corresponding to the
largest eigenvalue is defined as the DI of node $i$.
Computationally, DI can be easily calculated by power method.
Starting from a uniform vector $w^{(0)}=(1,1,\ldots,1)$, we mulitply
it with higher and higher powers of $M$ ($l$ is an integer):
\begin{equation}
w^{(l)}=(1,1,\ldots,1)M^l.\label{wl}
\end{equation}
Notice that, the $i$th entry of $w^{(l)}$ is the number of all
possible walks of length $l$ departing from node $i$. This explains
the idea of path counting behind the definition of DI.

This framework applies to a variety of dynamical models, including
epidemic spreading models, the Ising model
\cite{Brush1967,Dorogovtsev2002}, and diffusive processes like the
voter model \cite{Holley1975} or phase coupled oscillators
\cite{Acebron2005}. In the SIR model, DI is shown to be a good
predictor of spreading efficiency at the critical regime,
outperforming the predictions made by degree, $k$-shell index and
betweenness centrality.

Another approach based on path counting is a method to approximate
the number of infections resulting from a given initially-infected
node in a network of susceptible individuals \cite{Bauer2012}. This
method directly considers the spreading process and provides
estimation of actual number of infections with probability analysis.
The derivation of the impact of vertex $i$, i.e. the estimated
number of infections given that vertex $i$ was infected first, is a
little bit lengthy, so we will not introduce the details here. The
computation of node $i$'s impact requires the number of walks from
$i$ to all other nodes of arbitrary length $k$ with $l$ repeated
vertices, which is the bottleneck of the computation complexity for
this method. In reality, to avoid long computing time, the impact of
vertices can be estimated by imposing a maximal walk length $L$.
Simulations show that very good results can be obtained with short
path lengths $L=4$. We should note here that the exact definition of
impact of nodes in this method relies on the assumed spreading
models. So different models may result in different definitions.

Apart from the progresses mentioned above, a method measuring node
spreading power by expected cluster degree is proposed
\cite{Lawyer2012}. Authors quantify the spreading ability of a
source node by the expectation of the degree of a disease cluster
starting from it. The degree of a cluster of nodes is defined as the
number of edges that connect nodes within and outside the cluster.
Define the expected reach of node $i$, $ER_X(i)$, as the expectation
of the degree of the infected cluster after $X$ contagions starting
from $i$. In practice, the $ER_X(i)$ can be obtained by counting all
possible clusters of infected nodes which could appear after $X$
infections starting from $i$ and then taking the average cluster
degree. This method is applied to the recent Ebola outbreak in
Uganda, and it predicts that Ebola is unlikely to spread globally.
Similar with other path counting methods, computational expense is
still a major concern of $ER_X(i)$. This problem can be solved by
imposing a maximal $X$ value: it is shown $X=3$ is sufficient to
determine the outcome \cite{Lawyer2012}.

\section{Finding the most influential sets of nodes}

In real-world applications, there are many situations where we need
to find a small set of nodes that can spread information to the
largest number of nodes in the network. For example, in viral
marketing, a company tries to promote a new commercial product using
the word-of-mouth effects. The best strategy is to persuade more
people to buy the product with a limited advertising budget. In this
case, how to find the most influential sets of nodes that can lead
to maximal influence in complex networks becomes a fundamental
issue.

The problem of finding most influential multiple spreaders is
different from the one locating single influential spreaders. For
spreading processes originating from a set of nodes simultaneously,
the distance between these originators from each other should be
taken into consideration. This is because the nodes influenced by
the origins may have great overlap. Since the methods for single
influential spreaders are not guaranteed to find nodes that are far
enough \cite{Kitsak2010}, we cannot just select the top $k$ nodes
using the predictors designed for single super spreaders. Actually,
this issue has been abstracted as a fundamental algorithmic problem
in computer science \cite{Domingos2001}. For any spreading process,
there exists a influence function $f(\cdot)$ defined on a set of
nodes $S$. Assuming that $S$ is the set of originators, $f(S)$ is
the expected number of infected nodes at the end of spreading
process. With this interpretation, the problem becomes to maximize
$f(S)$ for the set $S$ with $k$ nodes. In fact, this is a rather
hard computational problem. Kempe et al. \cite{Kempe2003} has proved
that for a generic class of threshold models, it is NP-hard to find
the optimal set $S$. Therefore, what we can do is to find suboptimal
results with heuristic algorithms.

The basic idea behind the construction of heuristic algorithms lies
in the fact that, when we add nodes to the originators' set $S$,
usually the spreading will not increase significantly, but once the
right nodes are added, the process suddenly spreads widely. Such
property is described as submodularity mathematically
\cite{Kempe2003}. It has been shown most instances of threshold
models are submodular \cite{Kempe2003,Kempe2005}. For models with
such property, Kempe et al. \cite{Kempe2003} developed a greedy
search algorithm: starting from a randomly selected node, each time
we add the node that can lead to maximal increase of spreading if we
include it to $S$, until the desired size of $S$ is reached. Making
use of the classical theorem of Nemhauser, Wolsey, and Fisher
\cite{Nemhauser1978}, they proved analytically that the solution of
this greedy strategy is within $63\%$ of optimal for several classes
of models. This is the first provable approximation guarantees for
heuristic algorithms for this problem. However, the spread
estimation procedures, which are usually implemented by Monte Carlo
simulations, are quite time-consuming, thus limits the efficiency of
this algorithm.

Based on these fundamental results, several improvements have been
developed. To reduce the Monte Carlo simulations to estimate spread,
Leskovec et al. \cite{Leskovec2007a} exploited submodularity and
proposed a `Cost-Effective Lazy Forward` (CELF) optimization to the
simple greedy algorithm. The main idea is that the marginal gain of
a node in the current iteration cannot exceed its marginal gain in
previous iterations. With this idea, CELF optimization significantly
reduces the number of calls made to the spread estimation procedure.
Later on, Chen et al. \cite{Chen2009a} proposed an improved version
of original greedy algorithm, NewGreedy. More recently, Goyal et al.
\cite{Goyal2011} developed the algorithm CELF++, which is an
extension to CELF that further reduces the number of spread
estimation calls. Other approaches can be found in literature
\cite{Goyal2011a,Ben2009,Wang2010,Kimura2007,Kimura2009,Chen2010}.
These heuristic algorithms improve the computational efficiency of
the original greedy algorithms.

Besides the greedy strategies for submodular models, an efficient
algorithm based on message passing in statistical physics is
proposed recently \cite{Altarelli2013}. Message-passing algorithm is
an efficient method to deal with problems in statistical physics and
combinatorial optimization
\cite{Mezard2009,Altarelli2011,Altarelli2011a}. To solve the spread
optimization problem, authors mapped it on a high dimensional static
constraint-satisfaction model. Then they developed efficient
message-passing algorithms to find a solution to the spread
maximization problem. Compared with the greedy algorithm, this new
approach takes into account of the cooperative characteristics,
which are fundamental in real systems. With analytic and algorithmic
results on random graphs as well as a real-world network, it is
shown for a wide range of irreversible dynamics, even without
submodularity, the spreading optimization problem can be solved
efficiently on large networks.

\section{Empirical research on influential spreaders}

With the rapid development of the Internet and online communities, a
huge number of large scale datasets become available for researchers
to conduct analysis on spreading processes. In the last decade, huge
number of research works have been performed on datasets from
various types of online social networks, including email
communication \cite{Liben2008}, online social network - Facebook
\cite{Viswanath2009,Ahn2007,Mislove2007}, microblogging service -
Twitter
\cite{Weng2010,Krishnamurthy2008,Cha2010,Bakshy2011,Kwak2010,Gonzalez2011},
blogs sharing community - LiveJournal
\cite{Backstrom2006,Liben2005}, and other online communities
\cite{Adar2005,Gruhl2004,Centola2010,Goel2012,Aral2012,Muchnik2013,Sun2009,Bakshy2009,Aral2009}.

Directly examining the real diffusion data enables us to reveal the
exact spreading mechanisms beneath the propagation and discover new
principles dominating the diffusion process. In this section, we
would like to focus on the issue of finding individual super
spreaders and present our new results based on complete network
structure and the record the diffusion instances. As we have shown,
most of previous researches on individual super spreaders are based
on the assumptions of specific spreading rules. However, how the
information diffuses in reality is still not clear. Recently study
\cite{Goel2012} shows that the structure of online diffusion
networks cannot be fully described by current theoretical models.
And some empirical researches find that the prediction based on
specific models are not always correct
\cite{Centola2007,Goldenberg2001,Jackson2011,Aral2013}.
Consequently, we want to check the validity of these measures with
real diffusion data.

To achieve this, we have gathered the social network as well as
diffusion data from a well-known online blog community--LiveJournal.
This online community has been used to study spreading in previous
research works \cite{Liben2008,Backstrom2006,Liben2005}. The social
network is constructed by the friend relations in LiveJournal, i.e.
if user $a$ is in user $b$'s friend list, then there is a directed
social link from $b$ to $a$. In LJ, users can obtain the update of
the friends' posts. Therefore, the information would flow along the
incoming links of a source node. With this method, we obtain a
complete social network with 9,636,481 nodes and 197,368,009 links.
In order to extract the diffusion instances, we collect 56,180,137
posts published by LiveJournal users and filter 598,833 posts that
contain links to other posts in LiveJournal. In this way, if user
$a$ has cited user $b$'s posts at least once, we put a diffusion
link from $b$ to $a$. The resulting unweighted directed graph is
called diffusion graph, from which we can infer each node's
influence. In our study, we only consider the measures of indegree,
PageRank and $k$-shell. Notice that, according to the definition of
LJ social network, indegree describes the number of audiences of a
user. So here we discuss indegree rather than outdegree. Eigenvalue
centrality is similar to PageRank. And betweenness centrality is
infeasible for such large network due to its great computational
complexity. All the definitions of these measures can be found in
previous section.

\begin{figure}[ht]
\includegraphics[bb=0 0 1200 451,width=1\columnwidth]{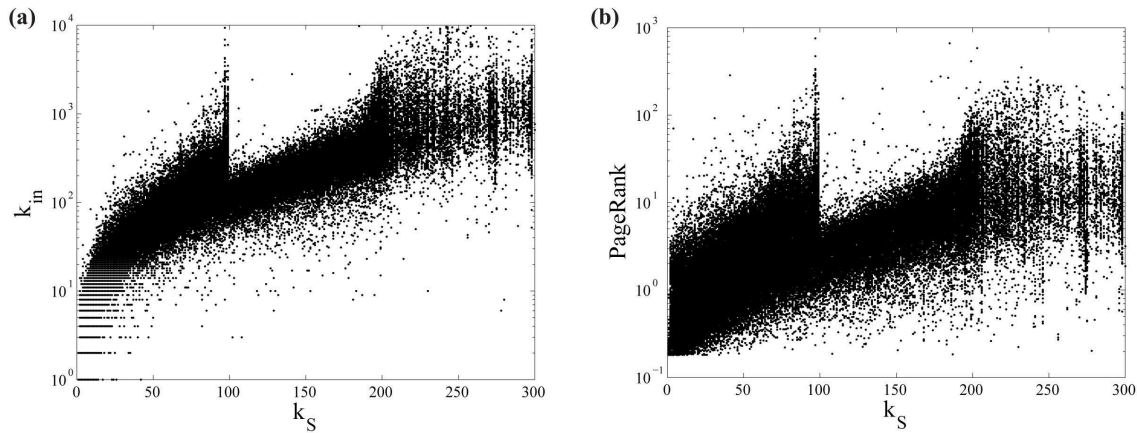}
\caption{(a), Indegree $k_{in}$ versus $k$-shell index $k_S$ for
nodes in LiveJournal. (b), Crosspolt of PageRank and $k_S$ for nodes
in LiveJournal. The correlations between $k_S$ and the other two
measures are weak.} \label{crossplot}
\end{figure}

Our first observation of this dataset is that, in the LiveJournal
social network with rich topological structure, different measures
can reflect very distinct information for each node. In
Fig.\ref{crossplot}(a), there exist lots of nodes with large
indegree but small $k_S$. Meanwhile, not all the nodes with large
$k_S$ have large indegree. The similar result is also obtain for
PageRank, which is presented in Fig.\ref{crossplot}(b). Since the
measures may convey different information for the same node, their
abilities to reflect spreading power should be different as well.

Despite the large number of users in LiveJournal, there are only a
small fraction of nodes participating in spreading. Precisely, only
246,423 users are involved in the information diffusion. To quantify
a node's spreading ability, we infer its influence by the size of
its outgoing component in the diffusion graph. Concretely, for each
node in diffusion graph, we first follow the diffusion links
starting from it, finding the first layer nodes that adopt the
information, and then track the links starting from these nodes and
so on. This process applies recursively until there are no more
diffusions exist. We define the number of these reachable nodes from
node $i$ in the diffusion graph as the influence of node $i$, and
denote it as $M_i$. Although the exact information diffusing from
node $i$ to these nodes may be different, node $i$ has great
potential to influence these nodes.

\begin{figure}[ht]
\includegraphics[bb=0 0 1200 902,width=1\columnwidth]{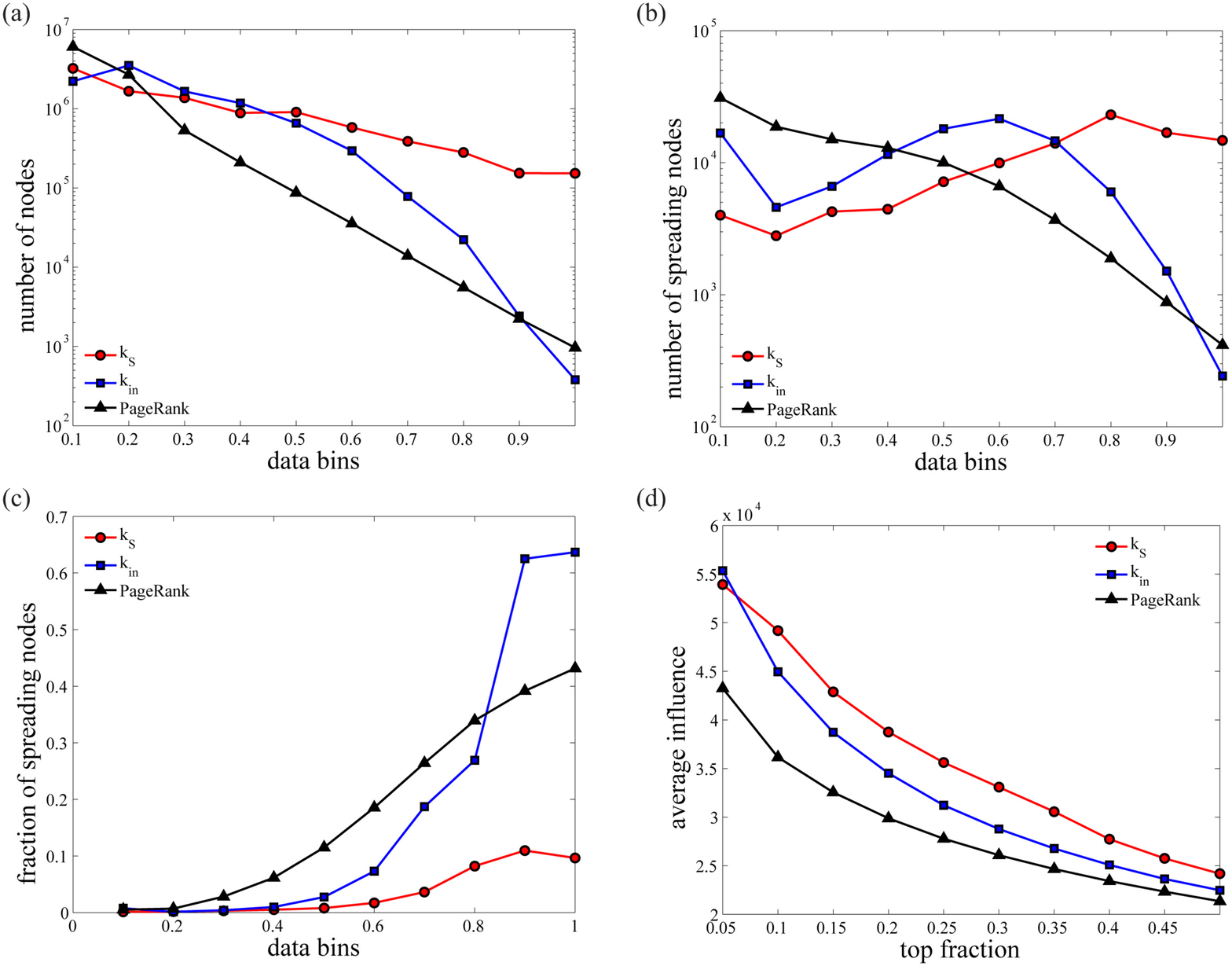}
\caption{(a), The number of users in each data bin for $k_S$,
$k_{in}$ and PageRank respectively. The bins are created by dividing
the range of each measure into ten parts equally according to the
logarithmic value. From 0 to 1, the measure increases from minimum
to maximum. (b), Number of nodes involved in spreading in each data
bin for $k_S$, $k_{in}$ and PageRank respectively. (c), Fraction of
users in each bin that participate in information diffusion. (d),
Average influence of users ranking top by different measures. We
rank the spreading users by different measures, select the nodes
ranking in top $f$ fraction, and then take the average of their
influence.} \label{compare}
\end{figure}

To compare the effect of different measures, in
Fig.\ref{compare}(a), we show the distribution of nodes for $k_S$,
$k_{in}$ and PageRank. To construct the data bins, we first take the
logarithmic values of each measure, and then divide the range into
ten intervals equally. Considering the different ranges of these
measures, the intervals are normalized to $[0,1]$, and from $0$ to
$1$, the value increases from the minimum to maximum. We can see in
the top region, $k_S$ has far more nodes than $k_{in}$ and PageRank.
This is because while $k_{in}$ and PageRank assign tens of thousands
different values, $k_S$ can only have several hundreds discrete
values. Fig.\ref{compare}(b) presents the number of users involved
in spreading in each bin. The main difference between $k_S$ and the
other two measures is that the number of spreading users has a
growing tendency as $k_S$ increases. The larger number of spreading
users in the top region of $k_S$ can be explained by the larger
number of users in this area, as shown in Fig.\ref{compare}(a).

In order to check the distribution of spreading users according to
different measures, we plot Fig.\ref{compare}(c) to show the
fraction of users participating in diffusion in each data bin for
three predictors. The fraction of spreading users is defined as the
ratio between the number of spreading users and the total number of
users in each bin. Clearly, the fraction of all three measures
increases as the value of each measure grows. This indicates that,
the users ranking top by these predictors have larger probability to
participating in information spreading. Particularly, in the top
area, nodes with large $k_{in}$ have much more chance to spread
information than $k_S$ and PageRank. The reason for that is there
are very few hubs with extremely large indegree. Even though the
number of spreading users are relatively small, when divided by the
number of hubs, the fraction becomes larger than that of $k_S$ and
PageRank.

While indegree can find nodes participating in information diffusion
with larger probability in the top region, it is desirable to check
the influence of these identified users. For each predictor, we
present the average influence of the nodes involved in diffusion
that rank in top $f$ fraction in Fig.\ref{compare}(d). The result
show that the nodes identified by $k_S$ in general have larger
influence than indegree and PageRank. This means that if a spreading
process starts in the core region of network, it will lead to larger
diffusion. This result coincides with the recent report of SIR and
SIS modeling on real-world networks \cite{Kitsak2010}. Therefore, in
practice identifying super individual spreaders with $k_S$ is more
reliable than indegree and PageRank.

\section{Conclusion and discussion}

Searching for influential spreaders in complex networks is a crucial
issue for many applications. In this paper, we make a review of the
most important theoretical models in describing spreading dynamics,
and introduce the current methods to identify both the individual
and multiple influential spreaders in various diffusion processes.
Through empirical diffusion data from LiveJournal, we find that in
practice different measures usually convey distinct information for
nodes in social networks. Of all the users in the network, only a
small fraction of users participate in spreading. Indegree can
locate nodes that involve in information diffusion with higher
probability than $k$-shell and PageRank. However, if we want to
identify nodes with large influence, it is preferred to use
$k$-shell index. Our results come from the direct analysis of
empirical diffusion data, thus providing practical instructions in
real-world applications.

Even though great improvement has been made in the research of
finding influential spreaders, there are still many problems we need
to investigate. For instance, our results of $k$-shell are only
tested on a specific online community LiveJournal. How the results
apply in other systems needs to be further examined. For both the
single and multiple spreaders, most of the current algorithms
require the topological structure of the underlying social network.
In contrast, it is usually difficult to reconstruct the social
network in practice. Consequently, some local algorithms are still
desirable to be developed. In the future research works, these
considerations would still attract attention from various domains
and lead to further exploration.

\section*{Acknowledgments}
We gratefully acknowledge funding by ARL under Cooperative Agreement
Number W911NF-09-2-0053, NIH and NSFC (11290141, 11201018). We thank
L. Muchnik for useful discussions and providing the LiveJournal
data.

\end{document}